\documentclass[journal,10pt,twocolumn]{IEEEtran}

\usepackage{amsmath,graphicx,amssymb,amsthm}

\usepackage{epstopdf}
\usepackage{multirow}

\usepackage[]{algorithmicx}
\usepackage{algpseudocode,algorithm}

\usepackage{cite}

\usepackage{color}
\usepackage{cuted}
\usepackage{flushend}
\usepackage[caption=false,font=normalsize,labelfont=sf,textfont=sf]{subfig}
\graphicspath{./figures/}

\newcommand{\subparagraph}{}
\usepackage[compact]{titlesec}
\titlespacing{\section}{0pt}{2ex}{1ex}
\titlespacing{\subsection}{0pt}{1ex}{0ex}
\titlespacing{\subsubsection}{0pt}{0.5ex}{0ex}
\begin{document}
	
	\title{Sparse Array Selection Across Arbitrary Sensor Geometries with Deep Transfer Learning}
	
	\author{Ahmet~M.~Elbir and Kumar Vijay~Mishra
		\thanks{A. M. E. is with the Department of Electrical and Electronics Engineering, Duzce University, Duzce, Turkey (e-mail: ahmetmelbir@gmail.com).}
		\thanks{K. V. M. is with The University of Iowa, Iowa City, IA 52242 USA (e-mail: kumarvijay-mishra@uiowa.edu).}
	}
	
	%
	
	\maketitle
	
	\begin{abstract}
		Sparse sensor array selection arises in many engineering applications, where it is imperative to obtain maximum spatial resolution from a limited number of array elements. Recent research shows that computational complexity of array selection is reduced by replacing the conventional optimization and greedy search methods with a deep learning network. However, in practice, sufficient and well-calibrated labeled training data are unavailable and, more so, for arbitrary array configurations. To address this, we adopt a deep transfer learning (TL) approach, wherein we train a deep convolutional neural network (CNN) with data of a source sensor array for which calibrated data are readily available and reuse this pre-trained CNN for a different, data-insufficient target array geometry to perform sparse array selection. Numerical experiments with uniform rectangular and circular arrays demonstrate enhanced performance of TL-CNN on the target model than the CNN trained with insufficient data from the same model. In particular, our TL framework provides approximately $20\%$ higher sensor selection accuracy and $10\%$ improvement in the direction-of-arrival estimation error.
	\end{abstract}
	
	
	\section{Introduction}
	Phased sensor arrays are now a widely adopted and pervasive technology, which finds applications in diverse areas such as radar, sonar, communications, acoustics, and ultrasound \cite{shenoy1994phased,frank2008advanced,herd2015evolution}. Their ability to steer the beam electronically without any mechanical motion provides high spatial selectivity and ability to adaptively reject interference. From the Nyquist-Shannon Theorem \cite{haupt2015timed}, the array must not admit less than two signal samples in a single spatial period $\lambda$, which is same as the operating wavelength of the array. Otherwise, spatial aliasing, i.e., or multiple main-lobes, show up in the beampattern leading to a reduced directivity. This affects the accuracy in estimating the bearings or directions-of-arrival (DoAs), that are known to be equivalent to spatial frequencies \cite{johnson1982application}, of unknown sources or targets. To avoid such effects, conventional phased sensor arrays feature elements that are uniformly spaced from each other. However, the complexity, size, and cost of such arrays becomes infeasible with the increase in the number of sensors. In this context, there is immense interest in thinned or sparse sensor arrays \cite{linebarger1993difference,haupt1994thinned}, which offer significantly reduced hardware.

	The number of elements in an array determine its degrees-of-freedom (DoFs), which, in turn, are related to the number of sources whose bearings can be ascertained by the sensor array. In particular, the DoFs of a full or filled sensor array with $N$ elements are achieved by a thinned array with only \(\mathcal{O}(\sqrt{N})\) sensors \cite{haupt1994thinned}. Further, when equipped with suitable parameter recovery algorithms \cite{dspCoprime1,mishra2017high,sedighi2019optimum}, thinned arrays yield negligible performance degradation and reduced mutual coupling \cite{boudaher2017mutual,superNestedArray,liu2017hourglass}. It has also been shown that sparse array beamforming algorithms yield similar spatial filtering performance as that of a full array \cite{elbirQuantizedCNN2019}.
	
	In general, searching for an optimal sparse sensor array is a combinatorial problem \cite{moffet1968minimum} whose computational complexity increases with the number of sensors. Since a closed-form solution is difficult to come by, several sub-optimal (although mathematically tractable) solutions have been proposed \cite{kozick1991linear,superNestedArray,sedighi2019optimum,antennaSelectionForMIMO,antennaSelectionViaCO, antennaSelectionKnapsack}. Lately, learning-based techniques have garnered much interest in sparse sensor communications \cite{elbirQuantizedCNN2019,elbir2019robust} and signal processing \cite{elbir2018cognitive,deepLearning4SignalProcessing}. In particular, deep learning (DL) has proven to be more computationally efficient than combinatorial search \cite{elbirQuantizedCNN2019,elbir2019deepursi}. 
	
	For an optimally sparse sensor array problem, \cite{elbir2018cognitive} proposed a deep learning approach in the context of a cognitive radar. It employed a deep convolutional neural network (CNN) trained with a large dataset of array outputs that optimizes the sensor placement to yield the lowest estimation error for DoA of radar's targets. This approach was later extended \cite{elbir2019deepursi,elbir2019robust} to sensor selection in massive multiple-output multiple-input (MIMO) communications. Later works combined this learning-based sparse subarray search with hybrid beamformer design \cite{elbirQuantizedCNN2019} and wideband channel estimation \cite{elbir2019deep,elbir2020low} in massive MIMO systems. 
	
	However, the CNN architectures in the aforementioned works are designed for a specific array geometry and are, therefore, inapplicable to different array configurations without significant re-training with new data. This arises from the assumption that the data used for training and testing are drawn from the same or similar distribution, which is difficult to guarantee in real world. In deep learning, this problem is called \textit{domain mismatch} \cite{duan2012domain}. On the other hand, labeling sufficient training data for all possible application domains is prohibitive. It has been shown \cite{kulis2011you} that it is possible to establish a reasonable model by exploiting the labeled data drawn from another sufficiently labeled \textit{source domain} which is closer to describing similar contents of the \textit{target domain}. This \textit{domain adaptation} (DA) enables knowledge transfer across domains. Lately, transfer learning (TL) has emerged as an effective domain adaptation technique, wherein the DL network learns domain-invariant models across source and target domains \cite{pan2010survey}, and has been applied to processing of image, 
	bio-medical, 
	radar, 
	and speech 
	signals.
	
	Apart from domain mismatch problem, networks such as CNN also suffer from a need of a large training database. When only a limited labeled data are available, CNN fails to optimally select the sensor subarrays. Since CNN objective functions are highly non-convex and convergence of optimization algorithms to a global optimum is not guaranteed, training with only large data could increase the probability of convergence. Alternatively, training forms such as convolutional autoencoders (CAE) \cite{vincent2010stacked} and TL are employed in data-limited applications. Some studies \cite{seyfiouglu2017deep} suggest that TL outperforms CAE especially when the sample sizes are very small.
	
	In this paper, we address the domain mismatch between various array geometries and lack of massive training data by developing a more efficient, deep TL-based sensor subarray selection approach. Indeed, sufficient datasets required to support the level of training CNNs need are unavailable, expensive or impossible to extract in many real-world sensor array applications. In this paper, we apply TL to enable the network in selecting sensor subarrays accurately even when limited labeled data are available. In particular, we transfer the features in the training data from one array geometry to a different array configuration. For example, we use a CNN trained with a uniform rectangular array (URA) to select sensors in a uniform circular array (UCA). This domain transfer is advantageous when a large off-line database is required for system identification or calibration \cite{comparisonOfMIMOArrays}. 
	
	Conventionally, DoA estimation across various geometries is performed with array transformation and array interpolation techniques \cite{ref_AI5}. However, for multiple targets and complex geometries, these techniques are difficult to come by \cite{rubsamen2008direction,liu2019doa}. Our approach is helpful in overcoming such limitations. In particular, we consider the sensor selection for DoA estimation, wherein we train a deep CNN to select the ``best'' subarray with lowest estimation error. We choose  Cram\'{e}r-Rao lower bound (CRB) as a metric to obtain the best subarray \cite{antennaSelectionKnapsack,elbir2018cognitive}. 
	Our TL-based strategy yields approximately $20\%$ improvement in sensor selection performance. Our extensive numerical experiments with different array geometries in both source and target domains demonstrate the effectiveness of our approach. Further, it demonstrates robustness against array imperfections induced by operating conditions;  earlier works have studied this problem in the context of autoencoders \cite{liu2018direction}.
	
	The rest of the paper is organized as follows. In the following section, we describe the system model and formulate the problem. In Section~\ref{sec:DL4AS}, we introduce our DL network design and apply deep TL to the same in Section~\ref{sec:DeepTransferLearning}. We validate our model with several numerical experiments in Section~\ref{sec:numexp} and conclude in Section~\ref{sec:summ}. Throughout this paper, we denote the vectors and matrices by boldface lower and upper case symbols, respectively. In case of a vector $\mathbf{a}$, $[\mathbf{a}]_{i}$ represents its $i$th element. For a matrix $\mathbf{A}$, $[\mathbf{A}]_{:,i}$ and $[\mathbf{A}]_{i,j}$ denote the $i$th column and the $(i,j)$-th entry, respectively. The $\mathbf{I}_N$ is the identity matrix of size $N\times N$; $\text{E}\left\lbrace \cdot \right\rbrace $, $\angle\{\cdot\}$, $\operatorname{\mathbb{R}e}\left\lbrace \cdot \right\rbrace$ and $\operatorname{\mathbb{I}m}\left\lbrace \cdot \right\rbrace$ designate the statistical expectation, phase, real and imaginary parts of the argument, respectively; $\textrm{Toeplitz}\{\cdot\}$ constructs a Toeplitz matrix with its vector argument; and $\odot$ denotes the point-wise (Hadamard) product. The combination of selecting $K$ terms out of $M$ is denoted by $\left( \begin{array}{c} M \\ K \end{array}\right) = \frac{M !}{K!(M-K)!}$. The notation expressing a convolutional layer with $N$ filters/channels of size $D\times D$, is given by  $N$@$ D\times D$.
	
	\section{System Model}
	\label{sec:ArrayModel}
	Consider an $M$-element sensor array receiving a signal $s(t_i)$ from the direction $\Theta=(\theta,\phi)$ where $\theta$ and $\phi$ are the elevation and azimuth angles of the source with respect to the sensor array, respectively. The received signal is narrowband and the source is in the far-field of the sensor array. Then, the output of the sensor array is \textcolor{black}{\cite{crbStoicaNehorai}}
	\begin{align}
	\label{signalModel}
	\mathbf{y}(t_i) = \mathbf{a}(\Theta)s(t_i) + \mathbf{n}(t_i),\hspace{10pt} 1\leq i\leq T,
	\end{align}
	where $T$ is the number of snapshots, $\mathbf{y}(t_i) = [y_1(t_i),\dots, y_M(t_i)]^T$ and ${y}_m(t_i)$ denotes the output of the $m$-th sensor for the $i$-th snapshot, $\mathbf{n}(t_i) = [n_1(t_i),\dots, n_M(t_i)]^T$ is the noise vector and $n_m(t_i)$ is zero-mean spatially and temporarily white Gaussian noise with variance $\sigma_n^2$,  $\mathbf{a}(\Theta) = [a_1(\Theta),\dots, a_M(\Theta)]^T$ is the $M\times 1$ steering vector. The $m$-th element of $\mathbf{a}(\Theta)$ is
	\begin{align}
	a_m(\Theta) = \exp\left\{-j\frac{2\pi}{\lambda}\mathbf{p}_m^T\mathbf{r}(\Theta)\right\},
	\end{align}
	where $\mathbf{r}(\Theta)$ depends on the source direction as
	\begin{align}
	\mathbf{r}(\Theta) = [\cos(\phi)\sin(\theta), \sin(\phi)\sin(\theta), \cos(\theta)]^T,
	\end{align}
	and $\mathbf{p}_m = [x_m,y_m,z_m]^T$ is the position of the $m$-th sensor in the Cartesian coordinate system. 
	
	In the context of sparse array selection, our goal is to choose the ``best" $K$ sensors in an $M$-element array in the sense that the lowest statistical mean-square-error (MSE), i.e., the CRB is achieved \cite{crbStoicaNehorai,friedlander}. Overall, $C = \left( \begin{array}{c} M \\ K \end{array}\right)  = \frac{M!}{K! (M-K)!}$ possible subarray choices are available. Therefore, we can treat sensor selection as a classification problem with $C$ classes. It seems impractical to visit all possible subarray configurations to arrive at the best subarray candidate. However, it has been shown \cite{elbir2018cognitive,elbirQuantizedCNN2019,elbirSampTA} that many subarray candidates yield the same CRB level because of the non-unique placement of sensors within the array. Hence, the distinct number of subarrays is very small. 
	Note that the literature suggests other statistical bounds \cite{performanceBoundsWWB} for DoA estimation but a closed-form solution of only CRB is available for higher dimensional arrays.
	
	We observe $\Theta$ as the inner product $\mathbf{p}_m^T\mathbf{r}(\Theta)$. The exponential form of $a_m(\Theta)$ suggests that this is a multi-dimensional spatial harmonic whose frequencies (and hence, DoAs) can be extracted through conventional as well as sparse reconstruction algorithms \cite{elbir2018cognitive,mishra2017high}. The uniqueness of spatial harmonic retrieval \cite{nion2010tensor} is directly related to the number of sensors in the array. For a URA of size $M_1\times M_2$, at least $M_1M_2 - \text{min}(M_1,M_2)$ sensors are required for a perfect DoA retrieval in a noiseless setting. Hence, in any sparse sensor array selection, $K$ must satisfy these guarantees.

	In our proposed TL framework, we first design a deep network to select the best subarrays in the source domain. Then, we transfer the sensor selection ability of this network to target domain. Here, we assume that the source domain is a larger data-set in comparison to the target domain. The deep network trained with source domain data performs better than the one with the target domain when limited data are available. 


	
	\section{DL Network Design For Sensor Selection}
	\label{sec:DL4AS}
	A DL network is defined as a non-linear mapping which categorizes and clusters the input data. Let $\mathcal{D} = \{\mathcal{D}^{(1)},\dots, \mathcal{D}^{(\textsf{D})}  \}$ and $\mathcal{Y} = \{\mathcal{Y}^{(1)},\dots, \mathcal{Y}^{(\textsf{Y})}\}$ denote the input and output labels for a dataset. Then, the deep classification network is represented as $\Sigma (\mathcal{D}) = \mathcal{Y}$ mapping the input data to the output labels which represent the best subarray indices. In the following, we present the details of input and output design of the deep network.
	\subsection{Input Data}
	\label{subsec:InputDat}
	The input to our DL network are the covariance matrices of the received signal. In particular, we use the real, imaginary and the phase information of the covariance matrix. Let $\mathbf{X}$ be an $M\times M\times 3$ real-valued matrix with $3$ "channel". Hence, we have $\mathcal{D}^{(i)} = \mathbf{X}$ for $i$-th input instant. Specifically, we define the $(i,j)$-th entry of the first and the second "channel" of the input data as $[\mathbf{X}_{(:,:,1)}]_{i,j} = \operatorname{Re}\{ [\mathbf{R}]_{i,j} \}$ and $[\mathbf{X}_{(:,:,2)}]_{i,j} = \operatorname{Im}\{ [\mathbf{R}]_{i,j} \}$, respectively. Similarly, the third "channel" is given by $[\mathbf{X}_{(:,:,3)}]_{i,j} = \angle\{ [\mathbf{R}]_{i,j} \}$. Although real and imaginary inputs are sufficient to describe the complex covariance matrix, feeding a third quantity such as phase (or magnitude) lets the network know that the first two inputs are related to each other.

	\subsection{Labeling}
	\label{subsec:labeling}
	We treat the sensor selection problem as a classification problem with $C$ classes. The class label comprises the positions of the sensor subarray corresponding to that class. Let $\mathcal{P}_c^{(k)} =  \{{x_k}^{(c)},{y_k}^{(c)},{z_k}^{(c)}  \}$ be the set of sensor coordinates in the $c$-th subarray for $k = 1,\dots,K$. Then the positions of the sensors for the $c$-th class form the set $\mathcal{Y}_c =\{\mathcal{P}_c^{(1)},\dots,\mathcal{P}_c^{(K)} \}$. Therefore, the set of all classes is $\mathcal{Y} = \{\mathcal{Y}_1,\mathcal{Y}_2,\dots,\mathcal{Y}_C\}$.
	
	In order to select the best subarrays in $\mathcal{Y}$, we compute the CRB for each element of $\mathcal{Y}$ as $c = 1,\dots,C$. Consider the $K\times 1$ subarray output
	\begin{align}
	\label{subarrayOutput}
	\mathbf{y}_c(t_i) =  \mathbf{a}_c(\Theta )s(t_i) + \mathbf{n}_c(t_i),
	\end{align}
	where $\mathbf{a}_c(\Theta)\in \mathbb{C}^{K}$ denotes the array steering vector corresponding to the subarray with position set $\mathcal{Y}_c$. Let $\mathbf{R}_c= \frac{1}{T}\sum_{i=1}^{T}\mathbf{y}_c(t_i)\mathbf{y}_c^H(t_i)$ be the $K\times K$ subarray sample covariance matrix for the $K\times 1$ subarray output $\mathbf{y}_c(t_i)$. We denote the partial derivatives of $\mathbf{a}_c(\Theta)$ with respect to $\theta$ and $\phi$ by $\dot{\mathbf{a}}_c(\theta) = \frac{\partial \mathbf{a}_c(\Theta)}{\partial_{\theta}}$ and  $\dot{\mathbf{a}}_{c}(\phi) = \frac{\partial \mathbf{a}_c(\Theta)}{\partial_{\phi}}$, respectively. The signal and noise variances are $\sigma_s^2$ and $\sigma_n^2 $, respectively.
	
	The CRBs for $\theta$ and $\phi$ in a single source scenario are \cite{crbStoicaNehorai}
	\begin{align}
	\kappa(\theta,\mathcal{Y}_c) = \frac{\sigma_n^2}{2T\operatorname{\mathbb{R}e}\bigg\{ \boldsymbol{\Pi}_{\theta} \odot (\sigma_s^4\mathbf{a}_c^H(\Theta)\mathbf{R}_c^{-1} \mathbf{a}_c(\Theta))\bigg\}},\label{CRB_Theta}\\
	\kappa(\phi,\mathcal{Y}_c) = \frac{\sigma_n^2}{2T\operatorname{\mathbb{R}e}\bigg\{ \boldsymbol{\Pi}_{\phi}\odot (\sigma_s^4\mathbf{a}_c^H(\Theta)\mathbf{R}_c^{-1} \mathbf{a}_c(\Theta))\bigg\}},\label{CRB_Phi}
	\end{align}
	where
	\begin{align}
	\boldsymbol{\Pi}_{\theta}  = \dot{\mathbf{a}}_{c}^H(\theta) \left[\mathbf{I}_K - \frac{\mathbf{a}_c(\Theta) \mathbf{a}_c^H(\Theta)}{K} \right]\dot{\mathbf{a}}_{c}(\phi),\\
	\boldsymbol{\Pi}_{\phi} = \dot{\mathbf{a}}_{c}^H(\phi) \left[\mathbf{I}_K - \frac{\mathbf{a}_c(\Theta) \mathbf{a}_c^H(\Theta)}{K} \right]\dot{\mathbf{a}}_c(\theta).
	\end{align}
	We define the absolute CRB \cite{ye2008two} for the directions $\Theta$ and $\mathcal{Y}_c$ as the root-mean-square value 
	\begin{align}
	\label{computeCRB}
	\kappa(\Theta,\mathcal{Y}_c) = \frac{1}{\sqrt{2}}[\kappa(\theta,\mathcal{Y}_c)^2 +\kappa(\phi,\mathcal{Y}_c)^2]^{1/2}.
	\end{align}
	For simplicity, we select $\sigma_s^2=1$ and define the signal to noise ratio in the training data as SNR$_{\text{TRAIN}} = 10\log_{10}(\sigma_s^2/\sigma_n^2)$.
	
	\begin{algorithm}[t!]
		\begin{algorithmic}[1]
			\caption{Training data generation.}
			\Statex {\textbf{Input:} \label{alg:TrainingDataGen.} Sensor positions $\{\mathbf{p}_m\}_{m=1}^M$, $K$, $T$, number of data realizations $L$, number of directions $P$ and SNR$_{\text{TRAIN}}$}.
			\Statex {\textbf{Output:} Training data $\mathcal{T}$ with dimensions $\{M\times M\times3 \times LP, LP\}$.}
			\State Generate $P$ DoA angles $\Theta_p = (\theta_p,\phi_p)$ for $p =1,\dots,P$.
			\State \textbf{for} $1\leq p \leq P$ \textbf{do}
			\State \indent \textbf{for} $1\leq l \leq L$ \textbf{do}
			\State \indent Generate the array output $\{ \mathbf{y}^{(l,p)}(t_i)\}_{i=1}^T$ as
			\begin{align}
			\mathbf{y}^{(l,p)}(t_i) = \mathbf{a}(\Theta_p)s^{(l,p)}(t_i) + \mathbf{n}^{(l,p)}(t_i), \nonumber
			\end{align}
			\indent for $s^{(l,p)}(t_i)\hspace{-1pt} \sim \mathcal{CN}(0,\sigma_s^2)$, $\mathbf{n}^{(l,p)}(t_i)\hspace{-2pt} \sim\hspace{-1pt}\mathcal{CN}(0,\sigma_n^2\mathbf{I})$.
			\State \indent Construct all $K\times 1$ subarray output configurations \par \indent $\mathbf{y}_c^{(l,p)}(t_i)$ as in (\ref{subarrayOutput}) from $\mathbf{y}^{(l,p)}(t_i)$ for $c = 1,\dots,C.$
			\State \indent Compute $\kappa(\Theta_p,\mathcal{Y}_c)$ for $c = 1,\dots,C$  by using the \par \indent covariance matrices ${\mathbf{R}}_c^{(l,p)}$.
			\State \indent Using $\kappa(\Theta_p,\mathcal{Y}_c)$, find the best subarray index as
			\par \indent  $\mathcal{B}_{\bar{c}}^{(l,p)}$ from (\ref{ReducedSet}).
			\State \indent Compute the full array covariance matrix $\mathbf{R}^{(l,p)}$ 
			\par \indent from  $\mathbf{y}^{(l,p)}(t_i)$, $i = 1,\dots,T$.
			\State \indent Construct the input data $\mathbf{X}^{(l,p)}$ as
			\begin{align}
			[\mathbf{X}_{(:,:,1)}^{(l,p)}]_{i,j}& = \operatorname{Re}\{ [\mathbf{R}^{(l,p)}]_{i,j} \},\nonumber \\
			[\mathbf{X}_{(:,:,2)}^{(l,p)}]_{i,j}& = \operatorname{Im}\{ [\mathbf{R}^{(l,p)}]_{i,j} \}, \nonumber\\
			[\mathbf{X}_{(:,:,3)}^{(l,p)}]_{i,j} &= \angle\{ [\mathbf{R}^{(l,p)}]_{i,j} \}.\nonumber
			\end{align}
			\State \indent Design the output label as $z^{(l,p)} = \mathcal{B}_{\bar{c}}^{(l,p)}$.
			\State \indent \textbf{end for} $l$
			\State \textbf{end for} $p$
			\State Construct training data by concatenating the input-output pairs: \noindent \small $ \mathcal{T} = \{ (\mathbf{X}^{(1,1)}, z^{(1,1)}), (\mathbf{X}^{(1,2)}, z^{(2,1)}),\dots, $ $(\mathbf{X}^{(1,L)}, z^{(L,1)}),$ $(\mathbf{X}^{(2,1)}, z^{(1,2)})\dots, (\mathbf{X}^{(P,L)}, z^{(L,P)})\}.$ \normalsize
		\end{algorithmic}
	\end{algorithm}
	
	Once $\kappa(\Theta,\mathcal{Y}_c)$ is computed for $c = 1,\dots,C$, the best subarray label $\mathcal{B}_{\bar{c}}$ is
	\begin{align}
	\label{ReducedSet}
	\mathcal{B}_{\bar{c}} = \arg \min_{c = 1,\dots,C }  \kappa(\Theta,\mathcal{Y}_c).
	\end{align}
	Here, the subscript $(\cdot)_{\bar{c}}$ denotes the index of best subarrays, $\bar{c} = 1,\dots, \bar{C}$, where $\bar{C}$ is the number of best subarrays. {\color{black}As $K$ increases, $C$ becomes very large. This makes the classification operation very difficult. However, experiments reveal that most of the sensor subarrays yield the same $\kappa(\Theta,\mathcal{Y}_c)$ because the non-unique sensor positions are common in many subarray combinations. Hence, $\bar{C} \ll C$ implying that only a handful of classes yield the lowest estimation errors \cite{elbir2018cognitive,elbirQuantizedCNN2019}. In Table~\ref{tableComparisonForNumberOfClasses}, we present the comparison of $C$ and $\bar{C}$ for a UCA with $M=16$ antennas. We note that $\bar{C}$ is very small, which leads an effective classification performance.}  After computing all best subarray indices, we finally construct the best subarray set as $\mathcal{B} = \{\mathcal{B}_{1},\dots,\mathcal{B}_{\bar{C}} \}$, where $\mathcal{B} \subset \mathcal{Y}$.

	\begin{table}[H]
		\caption{Number of classes $C$ and the reduced number of classes $\bar{C}$ for a UCA with $M=16$. \label{tableComparisonForNumberOfClasses}}
		\centering
		{\begin{tabular}{|c|c|c|c|c|c|c|}
				\hline
				&$K=3$ &$K=4$ &$K=5$ &$K=6$ &$K=7$ &$K=8$  \\
				\hline
				$C$& $560$ &$1820$ &$4368$ &$8008$ &$11440$ &$12870$  \\
				$\bar{C}$&$16$ &$10$ &$16$ &$11$ &$16$ &$16$  \\
				\hline
		\end{tabular}}{}
	\end{table}

	Algorithm~\ref{alg:TrainingDataGen.} lists the steps to generate the training data by incorporating the input and labels, as discussed above. The training data is then fed to the deep network represented by $\Sigma(\cdot): \mathbb{R}^{M\times M\times 3} \rightarrow \mathcal{Y}$ that maps the input data $\mathbf{X}$ to the corresponding class in $\mathcal{Y}$.

	%
	\begin{figure}[t]
		\centering
		{\includegraphics[width=.36\textheight]{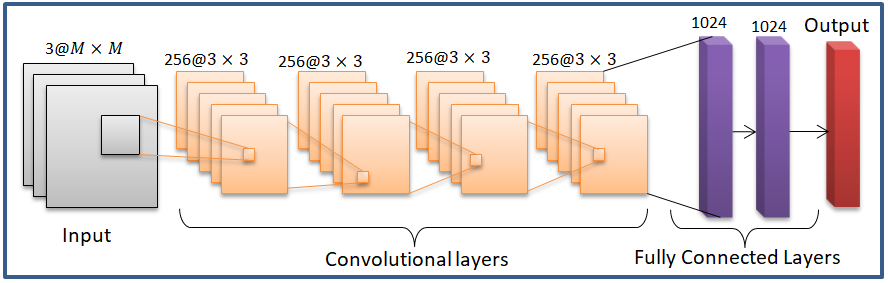}}
		\caption{The structure of the CNN for sensor selection. The middle layers that will be re-used from a pre-trained network for the target domain are shown in orange.}
		\label{figNetwork}
	\end{figure}
	
	\subsection{The Network Architecture}
	\label{subsec:NetworkArchitecture}
	Figure~\ref{figNetwork} illustrates the proposed deep network architecture for sensor selection. For multi-layer network, the non-linear function $\Sigma(\cdot)$ is represented by the inner layers as
	\begin{align}
	\Sigma(\mathcal{D}) = f^{(15)} \big(f^{(14)} ( \dots f^{(2)}(f^{(1)}( \mathcal{D} )   )   )   \big)  = \mathcal{Y},
	\end{align}
	where the first layer $f^{(1)}$ is the input layer and $f^{(i)}_{i \in\{2,4,6,8\}}$ denote the convolutional layers, each of which has 256 filters of size $3\times 3$. 	The arithmetic operation of a single filter of a  convolutional layer is defined for an arbitrary input  $\bar{\bf X} \in \mathbb{R}^{d_{x}\times d_{x}\times V_x}$ and output $\bar{\bf Y} \in \mathbb{R}^{d_{y}\times d_y\times V_{y}}$ as 
	\begin{align}
	\bar{\bf Y}_{p_y,v_y} = \sum_{p_k,p_x} \langle \bar{\bf W}_{v_y,p_k}, \bar{\bf X}_{p_x} \rangle,
	\end{align}
	where   $d_x \times d_y$ is the size of the convolutional kernel, $V_x \times V_y$ is the size of the response of a convolutional layer, $\bar{\bf W}_{v_y,v_k}\in \mathbb{R}^{V_x}$ denotes the weights of the $v_y$-th convolutional kernel, and $\bar{\bf X}_{p_x} \in \mathbb{R}^{V_x}$ is the input feature map at spatial position $p_x$. Hence, we define $p_x$ and $p_k$ as the two-dimensional (2-D) spatial positions in the feature maps and convolutional kernels, respectively \cite{quantizedCNN_Unified}.
	
	The $10$-th and $12$-th layer are fully connected with 1024 units whose $50\%$ is randomly selected during training to avoid overfitting. A fully connected layer maps an arbitrary input $\bar{\bf x}\in \mathbb{R}^{U_x}$ to the  output $\bar{\bf y}\in \mathbb{R}^{U_y}$ by using the weights  $\bar{\bf W} \in \mathbb{R}^{U_{x}\times U_{y}}$. Then, the $u_y$-th element of the output of the layer is the inner product
	\begin{align}
	\bar{\bf y}_{u_y} = \langle \bar{\bf W}_{u_y}, \bar{\bf x} \rangle = \sum_{i} {[\bar{\bf W}}]_{u_y,i}^\textsf{T} \bar{ \bf x}_i ,
	\end{align}
	for $u_y = 1,\dots, U_y$ and  $\bar{\bf W}_{u_y}$ is the $u_y$-th column vector of $\bar{\bf W}$, and $U_x = U_y = 1024$ is selected for $f^{(14)}$.
	
	After each convolutional and fully connected layers (i.e., $f^{(i)}_{i\in \{3,5,7,9,11,13\}}$), there is a rectified linear unit ($\mathrm{ReLU}$) layer  where $\mathrm{ReLU}(x) = \max(0,x)$. The $\mathrm{ReLU}$ layers are powerful in constructing the non-linearity of the deep network as well as providing non-negative output at the output layers, which is very useful for classification networks. The $14$-th layer has a classification layer with $\bar{C}$ units, where a $\mathrm{softmax}$ function is used to obtain the probability distribution of the classes. The  $\mathrm{softmax}$ layer is defined for an arbitrary input $\bar{\mathbf{x}}\in \mathbb{R}^{D}$ as $\mathrm{softmax}(\bar{{x}}_i) = \frac{\exp \{\bar{x}_i\} }{\sum_{i=1}^{D} \exp \{\bar{x}_i\} }$. The last layer $f^{(15)}$ is the classification layer.


	\section{Transfer Learning for Sensor Selection}
	\label{sec:DeepTransferLearning}
	\begin{figure}[t]
		\centering
		\subfloat[]{\includegraphics[scale=.53]{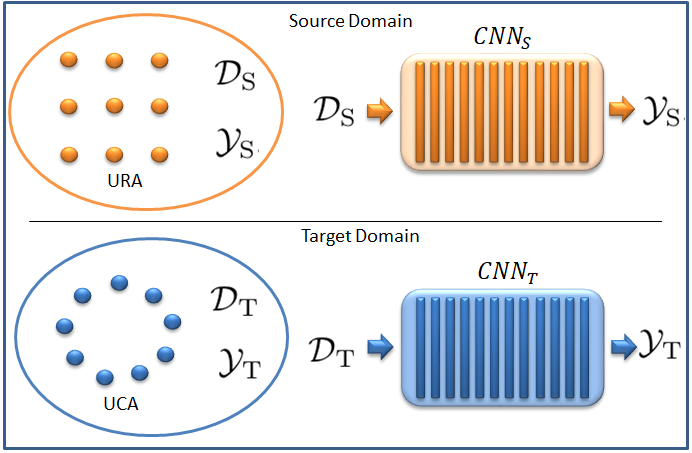}} \\
		\subfloat[]{\includegraphics[scale=.50]{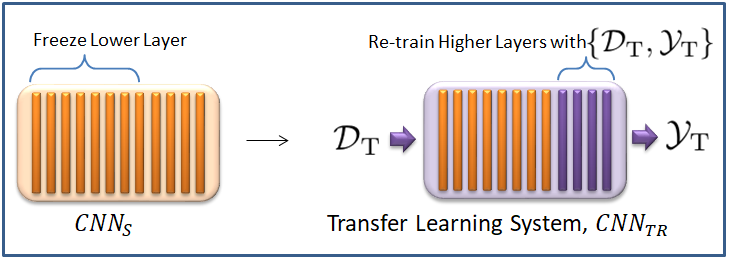}}
		\caption{The representation of the source and target domains for knowledge transfer from, for example, URA to UCA configuration. (a) Source (top) and target (bottom) domain data with corresponding learning networks $\mathrm{CNN}_\mathrm{S}$ and $\mathrm{CNN}_\mathrm{T}$, respectively. (b) In deep TL, lower layers of $\mathrm{CNN}_\mathrm{S}$ are frozen and only higher layers are re-trained with the target domain (UCA) data $\{\mathcal{D}_\mathrm{T}, \mathcal{Y}_\mathrm{T}\}$ to transfer sensor selection knowledge from source domain (URA).}
		\label{figTransNet}
	\end{figure}
	When compared with the domain transfer in shallow TL techniques \cite{pan2010domain}, such as classification based on support vector machine (SVM), a deep TL approach combines DA with the power of a deep network to learn the explanatory factors of variations in data and reduce the mismatch between the marginal distributions across array geometries. In Fig.~\ref{figTransNet}, we define the source (target) data and labels as $\mathcal{D}_\mathrm{S}$ ($\mathcal{D}_\mathrm{T}$) and $\mathcal{Y}_\mathrm{S}$ ($\mathcal{Y}_\mathrm{T}$), respectively. We train the source network $\mathrm{CNN}_\mathrm{S}$, which learns the non-linear relationship between $\mathcal{D}_\mathrm{S}$ and $\mathcal{Y}_\mathrm{S}$ as
	\begin{align}
	\label{Ys}
	\mathcal{Y}_S = \Sigma_\mathrm{S} (\mathcal{D}_\mathrm{S}),
	\end{align}
	where $\Sigma_\mathrm{S}(\cdot) $ is the non-linear function that constructs the mapping between the data and labels in the source domain. In (\ref{Ys}), the label data are the positions of the best subarray sensors as $\mathcal{Y}_\mathrm{S} = \{ \mathcal{Y}_\mathrm{S}^{(1)},\mathcal{Y}_\mathrm{S}^{(2)},\dots,\mathcal{Y}_\mathrm{S}^{(\textsf{S})} \}$ where $\textsf{S} = |\mathcal{Y}_\mathrm{S} | = |\mathcal{D}_\mathrm{S} |$ is the number of elements in the source domain.
	Furthermore, $\mathcal{D}_\mathrm{S}$ is the collection of covariance matrices of the array outputs of the source array geometry, i.e.,
	\begin{align}
	\mathcal{D}_\mathrm{S} = \{ \mathcal{D}_\mathrm{S}^{(1)},\mathcal{D}_\mathrm{S}^{(2)},\dots, \mathcal{D}_\mathrm{S}^{(\textsf{S})}   \},
	\end{align}
	where $\mathcal{D}_\mathrm{S}^{(i)} = \mathbf{X}_\mathrm{S}$ which is constructed from the source domain covariance matrix
	\begin{align}
	\mathbf{R}_\mathrm{S} = \frac{1}{T} \sum_{i = 1}^{T} \mathbf{y}_\mathrm{S} (t_i) \mathbf{y}_\mathrm{S}^H(t_i),
	\end{align}
	where $\mathbf{y}_\mathrm{S}(t_i)$ denotes the array output of the source data.
	
	Similarly, the target domain data and labels are $	\mathcal{D}_\mathrm{T} = \{ \mathcal{D}_\mathrm{T}^{(1)},\mathcal{D}_\mathrm{T}^{(2)},\dots, \mathcal{D}_\mathrm{T}^{(\textsf{T})}   \}$ and $	\mathcal{Y}_\mathrm{T} = \{ \mathcal{Y}_\mathrm{T}^{(1)},\mathcal{Y}_\mathrm{T}^{(2)},\dots, \mathcal{Y}_\mathrm{T}^{(\textsf{T})}   \}$, respectively, where $\textsf{T} = | \mathcal{D}_\mathrm{T}|$ and $	\mathbf{R}_\mathrm{T} = \frac{1}{T} \sum_{i = 1}^{T} \mathbf{y}_\mathrm{T} (t_i) \mathbf{y}_\mathrm{T}^H(t_i)$. For the target network $\mathrm{CNN}_\mathrm{T}$, we have
	\begin{align}
	\mathcal{Y}_\mathrm{T} = \Sigma_\mathrm{T} (\mathcal{D}_\mathrm{T}).
	\end{align}
	The TL framework assumes that the source domain has much larger dataset than the target domain, i.e., $\textsf{S} \gg \textsf{T}$. This implies that $\mathrm{CNN}_\mathrm{S}$ will turn out to be a well-trained deep network whereas $\mathrm{CNN}_\mathrm{T}$ has poor mapping performance and does not reflect the same mapping profile as $\mathrm{CNN}_\mathrm{S}$. To improve the performance of $\mathrm{CNN}_\mathrm{T}$, the key idea is to use the sensor selection ability of the pre-trained network $\mathrm{CNN}_\mathrm{S}$ even if it is trained with different array data~\cite{freeezingLayers3}. \textcolor{black}{This is achieved by re-training $\mathrm{CNN}_\mathrm{S}$ with the target domain data $\{\mathcal{D}_\mathrm{T},\mathcal{Y}_\mathrm{T}\}$ while freezing the lower layers (i.e., convolutional layers) of $\mathrm{CNN}_\mathrm{S}$\footnote{\color{black}We do not freeze the layers $\{3, 5, 7\}$, because they are $\mathrm{ReLU}$ layers with no weight to freeze.}. The new deep transfer network is $\mathrm{CNN}_\mathrm{TR}$ (Fig.~\ref{figTransNet}b).} The lower layers are kept intact or \textit{frozen} because they are generally domain invariant\footnote{``Domain invariance" implies that when new labels are added to the network, the lower layers remain unaffected even though the problem has changed.} and hence, harbor the bulk of sensor selection knowledge. The higher layers, however, are largely domain variant such that when new labels are added to the problem (i.e., $\mathcal{Y}_\mathrm{S}$ is replaced with $\mathcal{Y}_\mathrm{T}$), they require re-training. {\color{black}This approach  accelerates the computation of  the gradient in the backpropagation stage. Furthermore, it allows us to enlarge the feature space of the deep network without causing large error on the already-learned features~\cite{freeezingLayers3}.}
	
	\subsection{Knowledge Transfer Across Different Array Geometries}
	\label{knowledgeTransfer}
	Once $\mathrm{CNN}_\mathrm{S}$ (i.e., $\Sigma_\mathrm{S}(\cdot)$) is trained with the source domain data, we freeze the weights in the $\{2, 4, 6, 8\}$-th layers (i.e., the convolutional layers) to preserve the sensor selection ability of the deep network before transferring it to the target domain. We construct the TL network such that
	\begin{align}
	\Sigma_\mathrm{TR} (\mathcal{D}_\mathrm{T}) =  f^{(15)} \big(f^{(14)} ( \dots \tilde{f}^{(2)}(f^{(1)}( \mathcal{D}_\mathrm{T} )   )   )   \big)  = \mathcal{Y}_\mathrm{T},\label{TLmapping}
	\end{align}
	where the frozen layers are $\tilde{f}^{(i)}_{i \in \{2, 4, 6, 8\}}$. Algorithm~\ref{alg:TL} lists these steps of our proposed TL approach.

	\begin{algorithm}[t]
		\begin{algorithmic}[1]
			\caption{Transfer learning for sensor selection.}
			\Statex {\textbf{Input:} $\{\mathcal{D}_\mathrm{S}, \mathcal{Y}_\mathrm{S}\}$,  $\{\mathcal{D}_\mathrm{T}, \mathcal{Y}_\mathrm{T}\}$ }
			\label{alg:TL}
			\Statex {\textbf{Output:} $\mathrm{CNN}_\mathrm{TR}$.}
			\State Train $\mathrm{CNN}_\mathrm{S}$ with $\{\mathcal{D}_\mathrm{S}, \mathcal{Y}_\mathrm{S}\}$.
			\State Construct TL network $\mathrm{CNN}_\mathrm{TR}$ whose convolutional layers are designated the same as of $\mathrm{CNN}_\mathrm{S}$, i.e., ${f_\mathrm{TR}}^{(i)}_{i \in \{2, 4, 6, 8\}} = {f_\mathrm{S}}^{(i)}_{i \in \{2, 4, 6, 8\}}$.
			\State Train the remaining layers of the TL network with $\{\mathcal{D}_\mathrm{T}, \mathcal{Y}_\mathrm{T}\}$. Then, use $\mathrm{CNN}_\mathrm{TR}$ for sensor selection for target domain data.
		\end{algorithmic}
	\end{algorithm}
	
	\subsection{Deep Network Realization and Training}
	We realized the proposed TL architecture in MATLAB on a personal computer (PC) with 768-core graphics processing unit (GPU). For training, we used stochastic gradient descent algorithm with momentum $0.9$ and updated the network parameters at learning rate $0.01$ and mini-batch size of $512$. The loss function was the cross-entropy cost\par\noindent\small
	\begin{align}
	\label{costFunction}
	\mathrm{C}_\mathrm{E} = -\frac{1}{\bar{\textsf{T}}} \sum_{t = 1}^{\bar{\textsf{T}}} \sum_{c = 1}^{\bar{C}}  \bigg[\chi_c^{(t)} \ln \eta_c^{(t)} + (1 - \chi_c^{(t)}) \ln (1 - \eta_c^{(t)})  \bigg],
	\end{align}\normalsize
	where $\bar{\textsf{T}}$ is the length of the dataset and $\{\eta_c^{(t)}, \chi_c^{(t)}  \}_{t = 1, c=1}^{\bar{\textsf{T}}, \bar{C}}$ is the input-output pair for the classification layer. {\color{black}It is worth noting that the cost function in (\ref{costFunction}) can be defined in terms of the root-mean-square error (RMSE) of DoA estimation procedure. However, this makes the training process problem-dependent.} During training, the training data is shuffled for each epoch until training is terminated. Further, $80\%$ and $20\%$ of all generated data are chosen for training and validation datasets, respectively. The training rate is reduced by a factor of $0.9$ after each $10$ epochs. The training stops when the validation accuracy does not improve for three consecutive epochs.


	\section{Numerical Simulations}
	\label{sec:numexp}
	We validated the performance of our TL framework via several experiments. To train $\mathrm{CNN}_\mathrm{S}$, we collected array data for $P_\mathrm{S}=100$ equally spaced direction in the sector $\tilde{\Theta} = [0^\circ, 359^\circ]$ azimuth plane and $L_\mathrm{S}=100$ noisy data realizations with $T=100$ data snapshots. During training, we set $\sigma_s^2=1$ and use different SNR levels, namely, $\mathrm{SNR}_\mathrm{TRAIN} \in \{15, 20, 25\}$ dB. Hence, the total training data length is $3L_\mathrm{S}P_\mathrm{S}=30000$. Once $\mathrm{CNN}_\mathrm{S}$ is trained as outlined in Section~\ref{sec:DL4AS}, the $\mathrm{CNN}_\mathrm{TR}$ is constructed by following the steps in Algorithm~\ref{alg:TL}. For the above-mentioned settings with $M=16$ and $K=6$, the training time for $\mathrm{CNN}_\mathrm{S}$, $\mathrm{CNN}_\mathrm{T}$ are approximately $40$ and $5$ minutes respectively, whereas the TL network  $\mathrm{CNN}_\mathrm{TR}$ needs only $5$ seconds to be trained.
	
	\begin{figure}[t]
		\centering
		\includegraphics[width=.35\textheight]{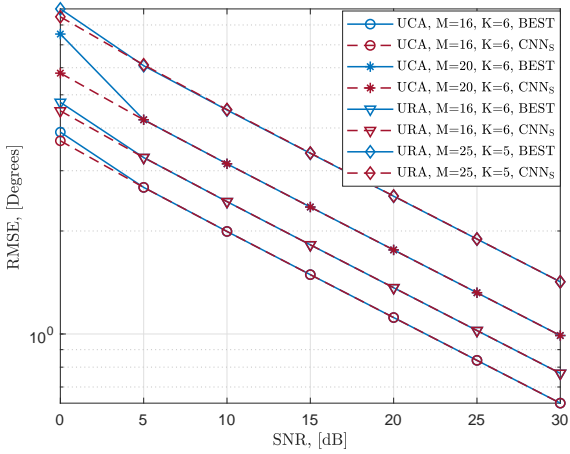} 
		\caption{DoA estimation performance for $\mathrm{CNN}_\mathrm{S}$ for  different array geometries.}
		\label{figSourceDOAAll}
	\end{figure}
	

	\begin{figure*}[t]
		\centering
		\subfloat[]{\includegraphics[width=.35\textheight]{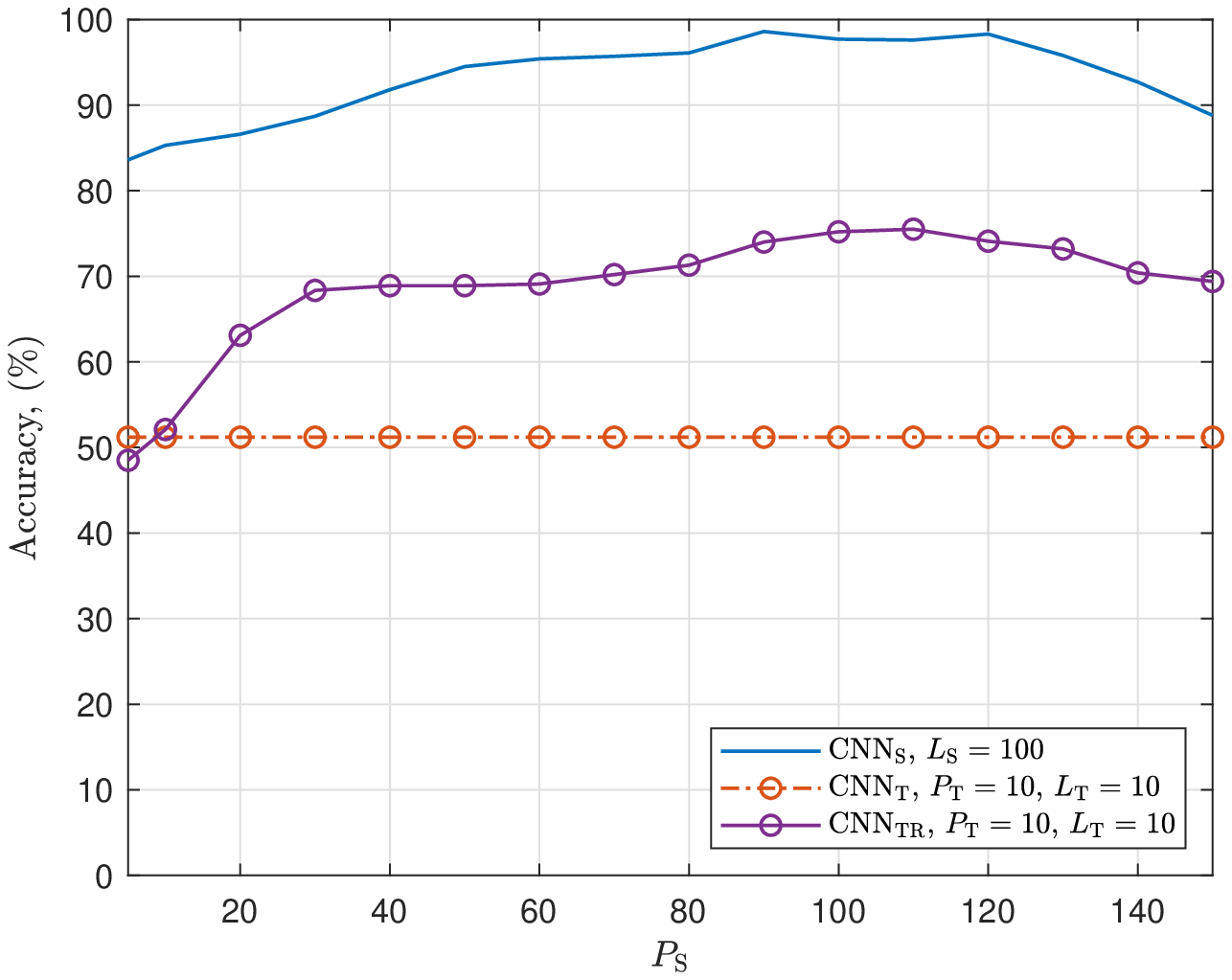}}
		\subfloat[]{\includegraphics[width=.35\textheight]{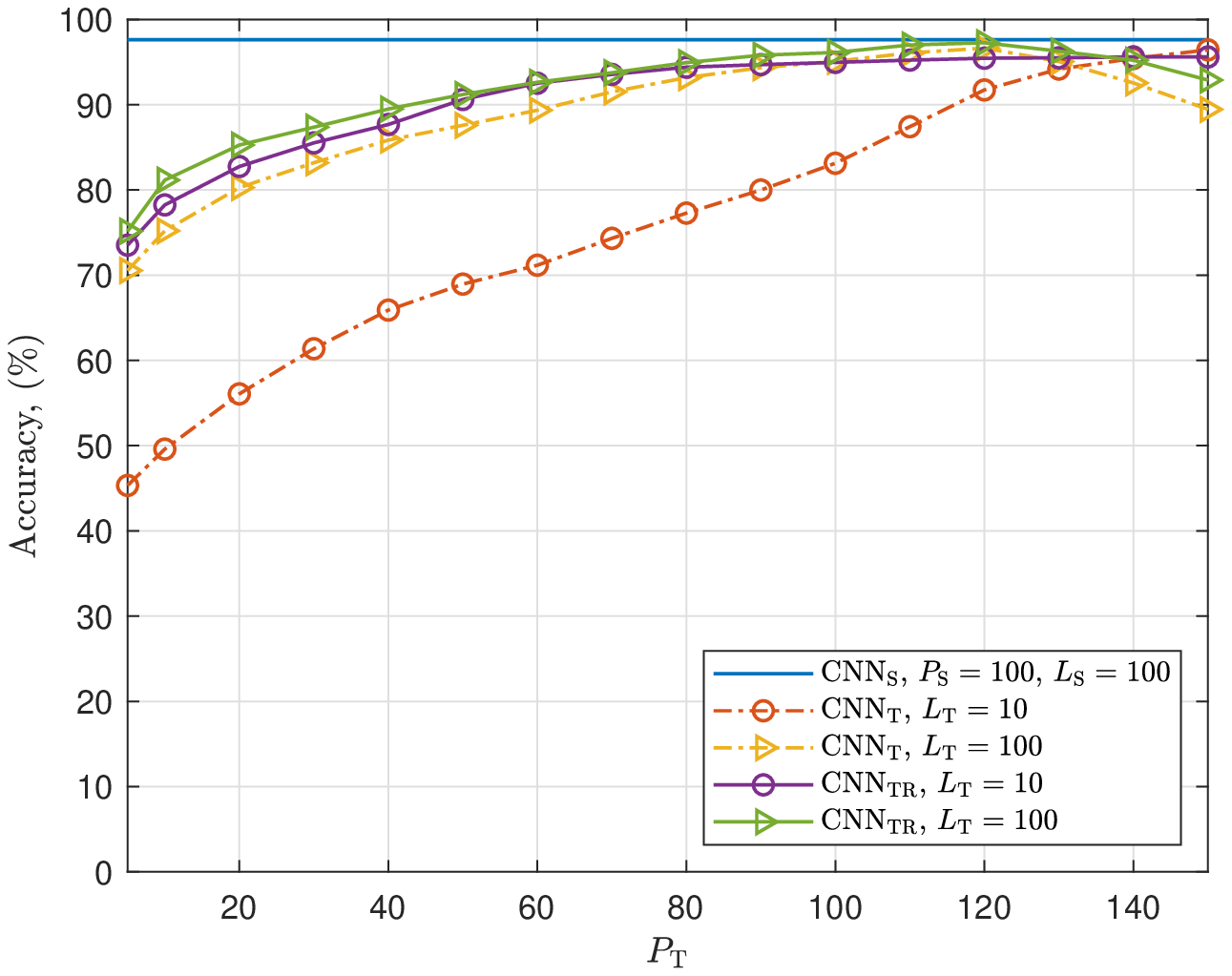}}
		\caption{\textcolor{black}{Performance of $\mathrm{CNN}_\mathrm{S}$, $\mathrm{CNN}_\mathrm{T}$ and $\mathrm{CNN}_\mathrm{TR}$  versus the number of DoA angles. Sensor selection accuracy is given with respect to (a) $P_\mathrm{S}$  when $L_\mathrm{S}=100$, $L_\mathrm{T}=10$, and (b) $P_\mathrm{T}$ when $P_\mathrm{S} = 100$, $L_\mathrm{S}=100$. $\mathrm{SNR}_\mathrm{{TRAIN}}=15$ dB.}}
		\label{figTLTest}
	\end{figure*}

	\subsection{Performance in Source Domain}	
	We first present the performance of the proposed CNN approach for the source domain case where different array geometries such as URA and UCA are considered with different array settings. In particular, we consider sensor arrays with half wavelength sensor spacing for both UCA and URA. When $\mathrm{CNN}_\mathrm{S}$s are trained for different arrays, we obtained above $90\%$ validation accuracy for the training data in all cases. In the prediction stage, the DoA angles are generated uniformly at random in the interval $\tilde{\Theta}$ so that the DoA angles in the training and prediction are selected from the same distribution. After feeding $\mathrm{CNN}_\mathrm{S}$ with these input data, the selected subarrays are obtained from the output for each scenario. Then, the sensor outputs of corresponding subarrays are employed for DoA estimation using MUSIC (MUltiple SIgnal Classification) algorithm~\cite{music}. During the simulations in the prediction state, the network is tested for different SNR levels for $J_T=100$ Monte Carlo trials. Figure~\ref{figSourceDOAAll} shows the RMSE in DoA estimation, i.e.,
	\begin{align}
	\mathrm{RMSE} = \bigg(\frac{1}{J_T  } \sum_{j=1}^{J_T} (\hat{\phi}^{(j)} - \phi)^2 \bigg)^{\frac{1}{2}},
	\end{align}
	where $\hat{\phi}^{(j)}$ and $\phi$ denote the estimated and true DoA angles, respectively. We compare the DoA estimation performance of $\mathrm{CNN}_\mathrm{S}$ with the best subarray that provides the lowest CRB. Figure~\ref{figSourceDOAAll} demonstrates that $\mathrm{CNN}_\mathrm{S}$ asymptotically follows the best subarray performance.

	\begin{table}[t]
		\caption{ Training Validation Accuracy (\%) For Different TL Scenarios}
		\label{tableValAcc}
		\centering
		\begin{tabular}{|c|c|c|}
			\hline
			\multirow{2}{*}{TL Scenario (Source $\rightarrow$ Target)}  & \multicolumn{2}{|c|}{Validation Accuracy (\%)} \\
			&         $\mathrm{CNN}_\mathrm{T}$    &  $\mathrm{CNN}_\mathrm{TR}$              \\
			\hline
			UCA $\rightarrow $ URA, $M=16$, $K=6$  &	54.9 & 	70.1	 \\
			\hline
			URA $\rightarrow $ UCA, $M=16$, $K=6$  &	42.3 & 	79.8	 \\
			\hline
			UCA $\rightarrow $ $\overline{\mathrm{UCA}}$, $M=20$, $K=6$  &	63.1 & 	98.8	 \\
			\hline
			URA $\rightarrow $ $\overline{\mathrm{URA}}$, $M=25$, $K=5$  &	55.2 & 	77.4	 \\
			\hline 
		\end{tabular}
	\end{table}
	
	\subsection{Performance for Transfer Learning}
	\label{sec:simTL}
	In order to evaluate the TL performance, we trained $\mathrm{CNN}_\mathrm{S}$ with different sizes of datasets and then constructed $\mathrm{CNN}_\mathrm{TR}$ from $\mathrm{CNN}_\mathrm{S}$ for sensor selection. We considered URA and UCA geometries with $M=16$, $K=6$ for source and target domains, respectively. {\color{black}Fig.~\ref{figTLTest} shows the sensor selection accuracy
		\begin{align}
		\mathrm{Accuracy}(\%) = \frac{\textsf{U}}{\textsf{V}}\times 100,
		\end{align} 
		where $\textsf{V}$ is the total number of input datasets in which the model identified the best subarrays correctly $\textsf{U}$ times. In Fig.~\ref{figTLTest}(a), the target domain $\mathcal{D}_\mathrm{T}$ are generated for $P_\mathrm{T}=10$ grid points in $\widetilde{\Theta}$ and $L_\mathrm{T}=10$ and we varied $P_\mathrm{S}$ from $5$ to $150$ for $\mathcal{D}_\mathrm{S}$ with $L_\mathrm{S}=100$. For all three networks,  The performance of $\mathrm{CNN}_\mathrm{T}$ is fixed because $\mathcal{D}_\mathrm{T}$ does not change during the simulations. When $P_\mathrm{S}$ is very small (i.e., $<10$), $\mathrm{CNN}_\mathrm{TR}$ performs even worse than $\mathrm{CNN}_\mathrm{T}$. However, as $P_\mathrm{S}$ increases, $\mathrm{CNN}_\mathrm{TR}$ and $\mathrm{CNN}_\mathrm{S}$ exhibit higher selection accuracy. For large source datasets, e.g. $P_\mathrm{S}\in [80, 120]$, $\mathrm{CNN}_\mathrm{TR}$ outperforms $\mathrm{CNN}_\mathrm{T}$ by a large margin because of the learned and transferred features from $\mathrm{CNN}_\mathrm{S}$. The increase in $P_\mathrm{S}$ does not necessarily improve the sensor selection performance because when the training data are densely sampled (i.e., $P_\mathrm{S}$ is high) the deep network cannot distinguish the input data of different directions and produce inaccurate classification output. These results suggest that $\mathrm{CNN}_\mathrm{S}$ needs to be trained with at least $\textsf{S} = P_\mathrm{S}L_\mathrm{S} = 40\cdot 100=4000$ to provide satisfactory accuracy (e.g., above $90\%$). As a result, $P_\mathrm{S} = 100$ is a reasonable choice for TL, wherein the target dataset $1000$ times smaller, i.e., $\frac{\textsf{S}}{\textsf{T}} = \frac{L_\mathrm{S}P_\mathrm{S}}{L_\mathrm{T}P_\mathrm{T}} = 1000$. In Fig.~\ref{figTLTest}(b), we repeat the same analysis for $\mathrm{CNN}_\mathrm{T}$ where we assume that $\mathrm{CNN}_\mathrm{S}$ is well-trained with $P_\mathrm{S}=100$ and $L_\mathrm{S}=100$. Then, we sweep $P_\mathrm{T}$ similarly for both $L_\mathrm{T}=10$ and $L_\mathrm{T}=100$. We can see that when $L_\mathrm{T} = 100$, $\mathrm{CNN}_\mathrm{T}$ quickly reaches maximum similar to $\mathrm{CNN}_\mathrm{S}$ as illustrated in Fig.~\ref{figTLTest}(a). In this case, the improvement gained by TL is incremental because $\mathrm{CNN}_\mathrm{T}$ is already well-trained. However, if small dataset is used, i.e., $L_\mathrm{T}=10$, then it requires larger $P_\mathrm{T}$ to reach high accuracy. Expectedly, this analysis shows that TL provides reasonable improvement if the target dataset is relatively small, i.e., $\textsf{T} = P_\mathrm{T}L_\mathrm{T}\leq 1000$ ($\textsf{S} = 10000$). In other words, when $\textsf{T}$ is high there is no need to use TL. Therefore, in the following experiments, we select $P_\mathrm{T}=L_\mathrm{T}=10$ and employ TL to improve the performance. }

	Table~\ref{tableValAcc} lists the validation accuracy of $\mathrm{CNN}_\mathrm{T}$ and $\mathrm{CNN}_\mathrm{TR}$ for different TL scenarios. We consider TL between UCA and URA as well as the perturbed array geometries denoted by $\overline{\mathrm{UCA}}$ and $\overline{\mathrm{URA}}$. In a perturbed array geometry, the $m$-th sensor position is selected uniformly at random as $\{\tilde{x}_m, \tilde{y}_m,\tilde{z}_m\} \sim \mathcal{N}(\{x_m, y_m, z_m\},(\lambda/4)^2)$ for each instance of the training data. It is evident that the sensor selection accuracy of $\mathrm{CNN}_\mathrm{TR}$ is approximately $20\%$ higher than $\mathrm{CNN}_\mathrm{T}$.
	
	\begin{figure}[t]
		\centering
		\includegraphics[width=.35\textheight]{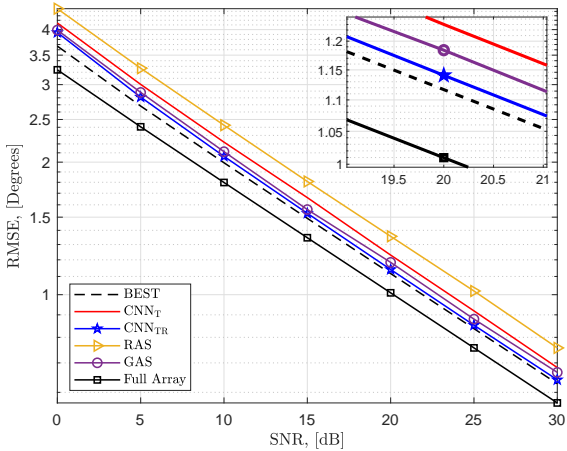} 
		\caption{DoA estimation performance when source domain has URA; target domain has UCA geometry. $M=16$, $K=6$.}
		\label{figTL_URA_UCA}
	\end{figure}
	\begin{figure}[t]
		\centering
		\includegraphics[width=.35\textheight]{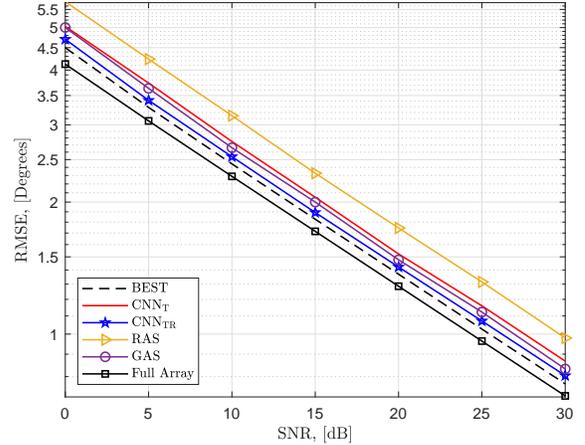} 
		\caption{DoA estimation performance when source domain has UCA; target domain has URA geometry. $M=16$, $K=6$.}
		\label{figTL_UCA_URA}
	\end{figure}

	We further assessed the DoA estimation performance of the selected subarrays 
	for target domain data. For $M=16$, $K=6$, Figs.~\ref{figTL_URA_UCA} and \ref{figTL_UCA_URA} depict the performance for URA$\rightarrow$UCA and UCA$\rightarrow$URA scenarios, respectively. We compared the sensor selection performance of $\mathrm{CNN}_\mathrm{T}$ and $\mathrm{CNN}_\mathrm{TR}$ {\color{black}with  greedy-based antenna selection (GAS)~\cite{antennaSelectionKnapsack}, random selection (RAS) as well as the fully array performance. As expected, we see that the full array has the lowest SNR due to large array aperture. We observe that $\mathrm{CNN}_\mathrm{TR}$ closely follows the performance of the best subarray. The $\mathrm{CNN}_\mathrm{TR}$ exhibits approximately $4\%$, $8.5\%$ and $23\%$ lower RMSE as compared to GAS, $\mathrm{CNN}_\mathrm{T}$ and RAS, respectively. It is worth noting that RAS has no rule on selecting the antennas while GAS is a greedy-based suboptimum method seeking the best subarray based on the CRB information~\cite{antennaSelectionKnapsack}.} These results establish the effectiveness of TL for DoA estimation with sensor selection. The superior performance of $\mathrm{CNN}_\mathrm{TR}$ is because of the learned and transferred features from source domain data via $\mathrm{CNN}_\mathrm{S}$.

	\subsection{Transfer Learning For Perturbed Sensor Positions}
\begin{figure}[t]
		\centering
		\includegraphics[width=.35\textheight]{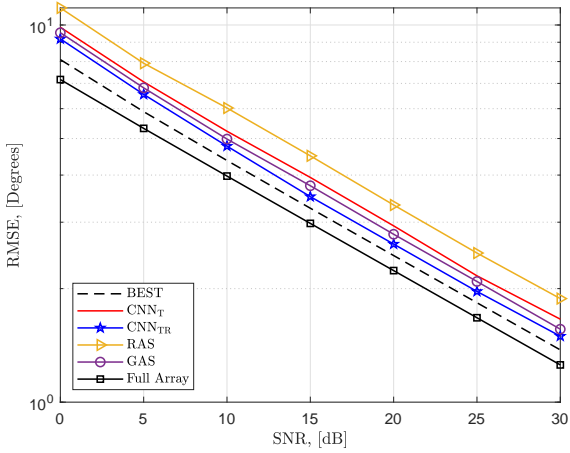} 
		\caption{DoA estimation performance when source domain has URA geometry and target domain has URA geometry with perturbed positions ($\overline{\mathrm{URA}}$), $M=25$, and $K=5$.}
		\label{figTL_URA_URAp}
	\end{figure}

\begin{figure}[t]
		\centering
		\includegraphics[width=.35\textheight]{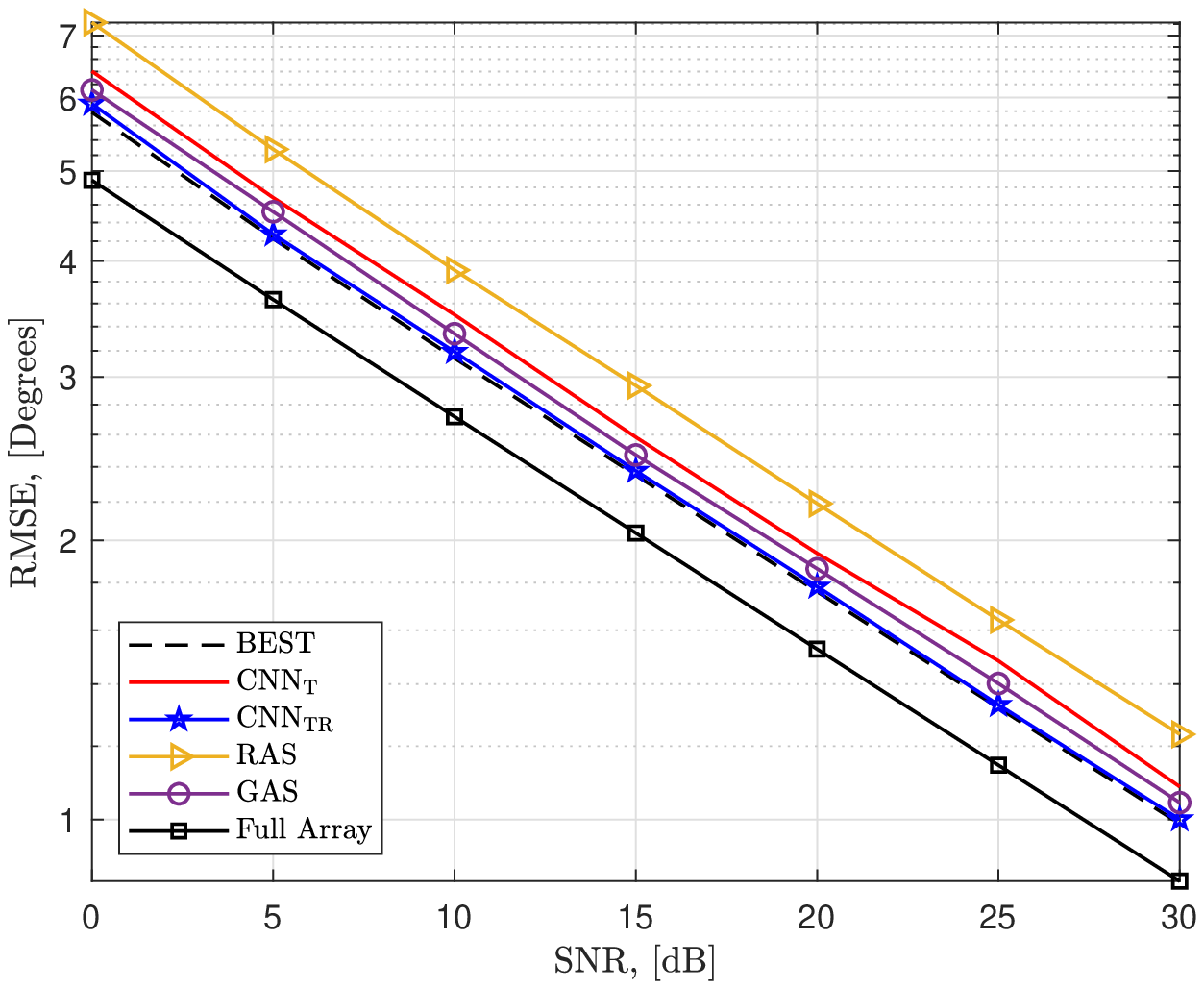} 
		\caption{DoA estimation performance when source domain has UCA geometry and target domain has UCA geometry with perturbed positions ($\overline{\mathrm{UCA}}$), $M=20$, and $K=6$.}
		\label{figTL_UCA_UCAp}
	\end{figure}

	In practical applications, the deployment of sensor arrays is a one-time operation. When the physical conditions around the sensor array change, the positions of the sensors are often \textit{slightly} altered. Over longer duration, the position of the sensors become different from the ones when the array is installed. In this experiment, we show that our TL approach for sensor arrays performs well even when the sensor positions are perturbed. Figures~\ref{figTL_URA_URAp} and \ref{figTL_UCA_UCAp} show the DoA estimation RMSE for URA and UCA, respectively. The target array geometry has been perturbed with $\lambda/4$ standard deviation in sensor positions (see Section~\ref{sec:simTL}). The proposed TL approach clearly results in lesser estimation error than $\mathrm{CNN}_\mathrm{T}$. In particular, $\mathrm{CNN}_\mathrm{TR}$ produces approximately {\color{black}$4\%$, $10\%$, and $28\%$ lower RMSE than GAS, $\mathrm{CNN}_\mathrm{T}$, and RAS, respectively.}

	{\color{black}
		\begin{figure}[t]
			\centering
			\includegraphics[width=.35\textheight]{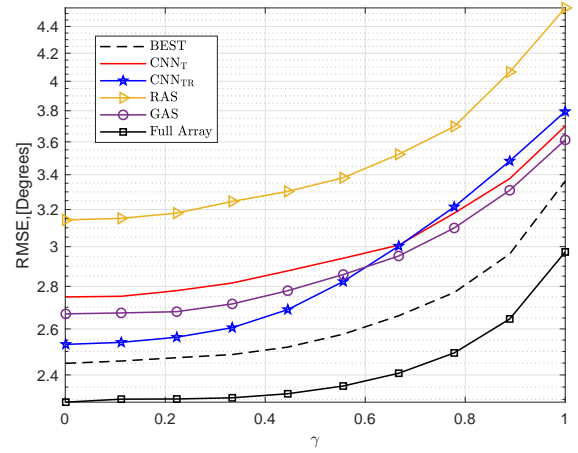} 
			\caption{DoA estimation performance vs. mutual coupling when source domain has URA; target domain has UCA geometry. $M=16$, $K=6$.}
			\label{figTL_Coupling_URA_UCA}
		\end{figure}
		\subsection{Transfer Learning For Sensor Data With Mutual Coupling}
		We assessed the performance of TL when the target data is corrupted. 
		We used the settings of Fig.~\ref{figTL_URA_UCA}, i.e., URA $\rightarrow $ UCA for $M=16$, $K=6$. The target sensor data is corrupted by mutual coupling (MC). The received signal now becomes~\cite{friedlander}
		\begin{align}
		\label{signalModelMC}
		\mathbf{y}(t_i) = \mathbf{C}\mathbf{a}(\Theta)s(t_i) + \mathbf{n}(t_i),\hspace{10pt} 1\leq i\leq T,
		\end{align}
		where $\mathbf{C}\in \mathbb{C}^{M\times M}$ is a Hermitian Toeplitz MC matrix, which for a UCA is $\mathbf{C} = \mathrm{Toeplitz}\{c_1,c_2,\dots,c_L, c_{L-1},\dots,c_2 \}$. Here $\{c_l\}_{l=1}^L$ are the MC coefficients and $L=\frac{M}{2} +1$ for even $M$. Let $\mathbf{c}=[c_1,\dots,c_L]^T$ be the MC coefficient vector, then we model $\mathbf{c}$ such that $c_1=1$ and $c_l = 0.6\big(1 - \frac{(l-2)}{L-1}\big)e^{j\varphi_l}$ for $l = 2,\dots,L$ where $\varphi_l\in [-\pi, \pi]$ is a random phase information. This yields that the magnitude of the coupling coefficient for the closest and furthest sensor pairs are $c_2 = 0.6$ and $c_L = 0.075$, respectively~\cite{elbirMCM}. To investigate the effect of MC, define $\mathbf{c} = \gamma [1/\gamma, c_2,\dots, c_L]^T$ and sweep $\gamma$ as $\gamma\in [0.01,1]$. The resulting performance in Fig.~\ref{figTL_Coupling_URA_UCA} shows that the performance of all algorithms degrades as $\gamma\rightarrow 1$, i.e., the effect of MC becomes stronger. $\mathrm{CNN}_\mathrm{TR}$ performs better than $\mathrm{CNN}_\mathrm{T}$ as long as $\gamma \leq 0.6$ because the corrupted data becomes unfamiliar to $\mathrm{CNN}_\mathrm{TR}$ and it yields worse RMSE than $\mathrm{CNN}_\mathrm{T}$ and GAS.
	}
	
	\subsection{2-D DoA Estimation}	
\begin{figure}[t]
		\centering
		\includegraphics[width=.33\textheight,height=.23\textheight]{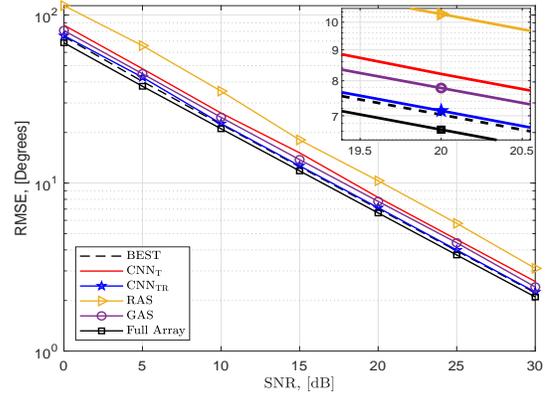} 
		\caption{2-D DOA estimation performance when source domain has UCA; target domain has URA geometry. $M=16$, $K_\mathrm{S}=6$ and $K_\mathrm{T}=8$.}
		\label{figTransferred2D} 
	\end{figure}

	So far, we restricted our experiments to a fixed elevation angle. Figure~\ref{figTransferred2D} shows the 2-D DoA estimation performance for the UCA$\rightarrow$URA scenario. In source domain, we selected $P_\mathrm{S} = 11000$ where the azimuth plane is sampled with $P_{{\phi}_\mathrm{S}}=100$ and the elevation plane is sampled uniformly with $P_{{\theta}_\mathrm{S}}=11$ in the sector $[80^{\circ},90^{\circ}]$. In target domain, we selected $P_\mathrm{T}=600$ ($\sim 5\%$ of $P_\mathrm{S}$) where $P_{{\theta}_\mathrm{T}}=6$ and $P_{{\phi}_\mathrm{T}}=10$. In this experiment, we consider different $K$ values, namely, $K_\mathrm{S}=6$ and $K_\mathrm{T}=8$. The number of snapshots are $T=10$. The RMSE is calculated for the joint estimation of $\theta$ and $\phi$. For 2-D scenario, {\color{black}the results are similar to the 1-D case: $\mathrm{CNN}_\mathrm{TR}$ has $7\%$, $20\%$, and $38\%$ lower RMSE than GAS, $\mathrm{CNN}_\mathrm{T}$, and RAS, respectively}. Note that the RMSE for all algorithms is high (approximately $70^\circ$ for $\mathrm{SNR} = 0$ dB) because of the small array aperture in vertical dimension \cite{elbirMCM}.

	{\color{black}
		\begin{table}[ht]
			\caption{Convolutional Layers Settings}
			\label{tableConvLayerParameters}
			\centering
			\begin{tabular}{|c|c|c|c|c|c|c|}
				\hline
				$l$&$D_x^{(l)}$& $D_y^{(l)}$ &$b_x^{(l)}$  & $b_y^{(l)}$  &   $N_\mathrm{CL}^{(l-1)}$  & $N_\mathrm{CL}^{(l)}$          \\
				\hline
				2 &	$M$ & $M$ &3	 & 3&	3 &256 \\
				\hline
				4  &	$M$ &  $M$ &3	 &3 &	256 &256 \\
				\hline
				6 &	$M$ & $M$ 	& 3 &3 &256	 &256\\
				\hline
				8 &	$M$ & 	$M$ 	&3 &3 &	256 &256  \\
				\hline 
			\end{tabular}
		\end{table}
		\begin{table}[ht]
			\caption{Fully Connected Layer Settings}
			\label{tableFCLayerParameters}
			\centering
			\begin{tabular}{|c|c|c|c|}
				\hline
				$l$&$D_1^{(l)}$& $D_2^{(l)}$    & $N_\mathrm{FCL}^{(l)}$        \\
				\hline
				10 &	$256 M^2$  	 &$1$&	$512$\\
				\hline
				12 &	$512$ & $1$&	$512$\\
				\hline 
			\end{tabular}
		\end{table}
		\subsection{Computational Complexity}
		Since all deep networks have the same architecture, $\mathrm{CNN}_\mathrm{S}$, $\mathrm{CNN}_\mathrm{T}$ and $\mathrm{CNN}_\mathrm{TR}$ have the same complexity. For a deep neural network with $L_\mathrm{C}$ convolutional layers, the time complexity is~\cite{vggRef}
		\begin{align}
		\mathcal{C}_\mathrm{CL} = 
		\mathcal{O}\bigg( \sum_{l=1}^{L_\mathrm{C}}D_x^{(l)}D_y^{(l)} b_x^{(l)}b_y^{(l)}N_\mathrm{CL}^{(l-1)}N_\mathrm{CL}^{(l)} \bigg),
		\end{align}
		where $D_x^{(l)}, D_y^{(l)}$ are the column and row sizes of each output feature map, $ b_x^{(l)},b_y^{(l)}$ are the 2D filter size of the $l$-th layer. $N_\mathrm{CL}^{(l-1)}$ and $N_\mathrm{CL}^{(l)}$ denote the number of input and output feature maps of the $l$-th layer respectively. In Table~\ref{tableConvLayerParameters}, we have shown the parameters of each convolutional layer. Thus, the complexity of $4$ convolutional layers with $256$@$3\times 3$ filters approximately becomes
		\begin{align}
		\mathcal{C}_\mathrm{CL} \approx \mathcal{O}\big(M^2 (4\cdot 9 \cdot 256^2) \big).
		\end{align}
		The time complexity of $L_\mathrm{F}$ fully connected layers similarly is
		\begin{align}
		\mathcal{C}_\mathrm{FCL}= \mathcal{O}\bigg(\sum_{l=1}^{L_\mathrm{F}}D_x^{(l)} D_y^{(l)} N_\mathrm{FCL}^{(l)}  \bigg), 
		\end{align}
		where $N_\mathrm{FCL}^{(l)}$ is the number of units of $l$-th fully connected layer and  $ D_1^{(l)}, D_2^{(l)}$ are the 2D input size of the $l$-th  fully connected layer and $N_\mathrm{FCL}^{(l)}$ is the number of units, each of which has $50\%$ dropout. Table~\ref{tableFCLayerParameters} lists the parameters of fully connected layers whose complexity approximately is 
		\begin{align}
		\mathcal{C}_\mathrm{FCL} \approx  \mathcal{O} \big(2\cdot 256^2 (M^2 + 2 )\big).
		\end{align}
		Hence the total time complexity of the DL approach is $\mathcal{C} = \mathcal{C}_\mathrm{CL} + \mathcal{C}_\mathrm{FCL}$ which is approximately
		\begin{align}
		\mathcal{C} \approx \mathcal{O}\big(M^2 (4\cdot 9 \cdot 256^2) +  2\cdot 256^2 (M^2 + 2)) \big),
		\end{align}
		which is further simplified as $ \approx \mathcal{O}\big(38\cdot 256^2 M^2  \big)$. In comparison, the order of an analytical approach such as GAS is $\mathcal{O}\big( KM^2 \big)$~\cite{antennaSelectionKnapsack}. The RAS has sorting complexity of $\mathcal{O}\big( M\log M\big)$ \cite{antennaSelectionViaCO} at the cost of performance. While the complexity of CNN is on the order of magnitude of $10^6$, it is able to run in more efficient parallel manner by using GPUs, whereas the other algorithms cannot be implemented in such a way easily. The computation time of the proposed CNN approach only takes about $10 \times 10^{-3}$ s for $M=16$ and $K=6$, whereas RAS and GAS need approximately $30\times 10^{-3}$ and $130\times 10^{-3}$, respectively. Similar observations about the fast computation times of the DL networks have been reported in~\cite{vggRef,deepCNN_ChannelEstimation,elbir2018cognitive,elbir2020DeepMUSIC}.
		
	}

	\section{Summary}
	\label{sec:summ}
	We proposed a deep TL framework for sparse sensor selection. We transfer the learned features from one domain of larger data length to another domain where limited number of observations are available. This is especially suitable for sensor placement applications where diverse geometries of arrays are encountered. Our deep TL approach provides significant performance improvement for sensor selection and DoA estimation for both uniform and non-uniform array geometries. 
	Moreover, TL is also effective for perturbed array geometries. This property allows us to first train a deep network with array data when it is deployed in field operations. When environmental and operational factors lead to deviations in the sensor positions, our approach is effective in overcoming the subsequent performance loss in DoA estimation.	In particular, our TL framework provides approximately $20\%$ more sensor selection accuracy and $10\%$ improvement in the DoA estimation RMSE.
	
	\bibliographystyle{ieeetr}
	\bibliography{IEEEabrv,references_036}

\begin{thebibliography}{10}

\bibitem{shenoy1994phased}
R.~P. Shenoy, ``Phased array antennas,'' in {\em Advanced radar techniques and
  systems} (G.~Galati, ed.), Peter Peregrinus, 1993.

\bibitem{frank2008advanced}
J.~Frank and J.~D. Richards, ``Phased array radar antennas,'' in {\em Radar
  handbook} (M.~I. Skolnik, ed.), McGraw-Hill Education, third~ed., 2008.

\bibitem{herd2015evolution}
J.~S. Herd and M.~D. Conway, ``The evolution to modern phased array
  architectures,'' {\em Proceedings of the IEEE}, vol.~104, no.~3,
  pp.~519--529, 2015.

\bibitem{haupt2015timed}
R.~L. Haupt, {\em Timed Arrays: Wideband and Time Varying Antenna Arrays}.
\newblock John Wiley \& Sons, 2015.

\bibitem{johnson1982application}
D.~H. Johnson, ``The application of spectral estimation methods to bearing
  estimation problems,'' {\em Proceedings of the IEEE}, vol.~70, no.~9,
  pp.~1018--1028, 1982.

\bibitem{linebarger1993difference}
D.~A. Linebarger, I.~H. Sudborough, and I.~G. Tollis, ``Difference bases and
  sparse sensor arrays,'' {\em IEEE Transactions on information theory},
  vol.~39, no.~2, pp.~716--721, 1993.

\bibitem{haupt1994thinned}
R.~L. Haupt, ``Thinned arrays using genetic algorithms,'' {\em IEEE
  Transactions on Antennas and Propagation}, vol.~42, no.~7, pp.~993--999,
  1994.

\bibitem{dspCoprime1}
X.~Wang, Z.~Chen, S.~Ren, and S.~Cao, ``{DOA} estimation based on the
  difference and sum coarray for coprime arrays,'' {\em Digital Signal
  Processing}, vol.~69, pp.~22 -- 31, 2017.

\bibitem{mishra2017high}
K.~V. Mishra, I.~Kahane, A.~Kaufmann, and Y.~C. Eldar, ``High spatial
  resolution radar using thinned arrays,'' in {\em IEEE Radar Conference},
  pp.~1119--1124, 2017.

\bibitem{sedighi2019optimum}
S.~S., M.~R. Bhavani~Shankar, K.~V. Mishra, and B.~Ottersten, ``Optimum design
  for sparse {FDA-MIMO} automotive radar,'' in {\em Asilomar Conference on
  Signals, Systems, and Computers}, 2019.
\newblock in press.

\bibitem{boudaher2017mutual}
E.~BouDaher, F.~Ahmad, M.~G. Amin, and A.~Hoorfar, ``Mutual coupling effect and
  compensation in non-uniform arrays for direction-of-arrival estimation,''
  {\em Digital Signal Processing}, vol.~61, pp.~3--14, 2017.

\bibitem{superNestedArray}
C.~L. Liu and P.~P. Vaidyanathan, ``{Super Nested Arrays: Linear Sparse Arrays
  With Reduced Mutual Coupling; Part I: Fundamentals},'' {\em IEEE Transactions
  on Signal Processing}, vol.~64, pp.~3997--4012, Aug 2016.

\bibitem{liu2017hourglass}
C.-L. Liu and P.~P. Vaidyanathan, ``Hourglass arrays and other novel {2-D}
  sparse arrays with reduced mutual coupling,'' {\em IEEE Transactions on
  Signal Processing}, vol.~65, no.~13, pp.~3369--3383, 2017.

\bibitem{elbirQuantizedCNN2019}
A.~M. {Elbir} and K.~V. {Mishra}, ``Joint antenna selection and hybrid
  beamformer design using unquantized and quantized deep learning networks,''
  {\em {IEEE} Trans. Wireless Commun.}, vol.~19, no.~3, pp.~1677--1688, 2020.

\bibitem{moffet1968minimum}
A.~Moffet, ``Minimum-redundancy linear arrays,'' {\em IEEE Transactions on
  antennas and propagation}, vol.~16, no.~2, pp.~172--175, 1968.

\bibitem{kozick1991linear}
R.~J. Kozick and S.~A. Kassam, ``Linear imaging with sensor arrays on convex
  polygonal boundaries,'' {\em IEEE Transactions on Systems, Man, and
  Cybernetics}, vol.~21, no.~5, pp.~1155--1166, 1991.

\bibitem{antennaSelectionForMIMO}
T.~M. Duman and A.~Ghrayeb, ``Antenna selection for {MIMO} systems,'' in {\em
  Coding for MIMO Communication Systems}, pp.~287--315, John Wiley \& Sons,
  2007.

\bibitem{antennaSelectionViaCO}
S.~Joshi and S.~Boyd, ``Sensor selection via convex optimization,'' {\em IEEE
  Transactions on Signal Processing}, vol.~57, no.~2, pp.~451--462, 2009.

\bibitem{antennaSelectionKnapsack}
H.~Godrich, A.~P. Petropulu, and H.~V. Poor, ``Sensor selection in distributed
  multiple-radar architectures for localization: {A} knapsack problem
  formulation,'' {\em IEEE Transactions on Signal Processing}, vol.~60, no.~1,
  pp.~247--260, 2012.

\bibitem{elbir2019robust}
A.~M. Elbir and K.~V. Mishra, ``Robust hybrid beamforming with quantized deep
  neural networks,'' in {\em IEEE International Workshop on Machine Learning
  for Signal Processing}, pp.~1--6, 2019.

\bibitem{elbir2018cognitive}
A.~M. Elbir, K.~V. Mishra, and Y.~C. Eldar, ``Cognitive radar antenna selection
  via deep learning,'' {\em IET Radar, Sonar \& Navigation}, vol.~13,
  pp.~871--880, 2019.

\bibitem{deepLearning4SignalProcessing}
D.~Yu and L.~Deng, ``Deep learning and its applications to signal and
  information processing [exploratory dsp],'' {\em IEEE Signal Processing
  Magazine}, vol.~28, no.~1, pp.~145--154, 2011.

\bibitem{elbir2019deepursi}
A.~M. Elbir and K.~V. Mishra, ``Deep learning design for joint antenna
  selection and hybrid beamforming in massive {MIMO},'' in {\em IEEE
  International Symposium on Antennas and Propagation and USNC-URSI Radio
  Science Meeting}, pp.~1585--1586, 2019.

\bibitem{elbir2019deep}
A.~M. Elbir and K.~V. Mishra, ``Online and offline deep learning strategies for
  channel estimation and hybrid beamforming in multi-carrier mm-{W}ave massive
  {MIMO} systems,'' {\em arXiv preprint arXiv:1912.10036v2}, 2020.

\bibitem{elbir2020low}
A.~M. Elbir and K.~V. Mishra, ``Low-complexity limited-feedback deep hybrid
  beamforming for broadband massive {MIMO} communications,'' in {\em IEEE
  International Workshop on Signal Processing Advances in Wireless
  Communications}, 2020.
\newblock in press.

\bibitem{duan2012domain}
L.~Duan, I.~W. Tsang, and D.~Xu, ``Domain transfer multiple kernel learning,''
  {\em IEEE Transactions on Pattern Analysis and Machine Intelligence},
  vol.~34, no.~3, pp.~465--479, 2012.

\bibitem{kulis2011you}
B.~Kulis, K.~Saenko, and T.~Darrell, ``What you saw is not what you get:
  {D}omain adaptation using asymmetric kernel transforms,'' in {\em IEEE
  Conference on Computer Vision and Pattern Recognition}, pp.~1785--1792, 2011.

\bibitem{pan2010survey}
S.~J. Pan, Q.~Yang, {\em et~al.}, ``A survey on transfer learning,'' {\em IEEE
  Transactions on Knowledge and Data Engineering}, vol.~22, no.~10,
  pp.~1345--1359, 2010.

\bibitem{vincent2010stacked}
P.~Vincent, H.~Larochelle, I.~Lajoie, Y.~Bengio, and P.-A. Manzagol, ``Stacked
  denoising autoencoders: {L}earning useful representations in a deep network
  with a local denoising criterion,'' {\em Journal of machine learning
  research}, vol.~11, no.~Dec, pp.~3371--3408, 2010.

\bibitem{seyfiouglu2017deep}
M.~S. Seyfio{\u{g}}lu and S.~Z. G{\"u}rb{\"u}z, ``Deep neural network
  initialization methods for micro-{D}oppler classification with low training
  sample support,'' {\em IEEE Geoscience and Remote Sensing Letters}, vol.~14,
  no.~12, pp.~2462--2466, 2017.

\bibitem{comparisonOfMIMOArrays}
B.~Chen, Z.~Zhong, B.~Ai, and X.~Chen, ``Comparison of antenna arrays for mimo
  system in high speed mobile scenarios,'' in {\em 2011 IEEE 73rd Vehicular
  Technology Conference (VTC Spring)}, pp.~1--5, 2011.

\bibitem{ref_AI5}
P.~Hyberg, M.~Jansson, and B.~Ottersten, ``Array interpolation and bias
  reduction,'' {\em IEEE Transactions on Signal Processing}, vol.~52, no.~10,
  pp.~2711--2720, 2004.

\bibitem{rubsamen2008direction}
M.~Rubsamen and A.~B. Gershman, ``Direction-of-arrival estimation for
  nonuniform sensor arrays: {F}rom manifold separation to {F}ourier domain
  {MUSIC} methods,'' {\em IEEE Transactions on Signal Processing}, vol.~57,
  no.~2, pp.~588--599, 2008.

\bibitem{liu2019doa}
Y.~Liu, H.~Chen, Z.~Peng, and J.~Fang, ``{DOA} estimation for mixed circular
  and noncircular signals by using the conversion relationship between {URA}s
  and a virtual {ULA},'' {\em IEEE Sensors Letters}, vol.~3, no.~11, pp.~1--4,
  2019.

\bibitem{liu2018direction}
Z.-M. Liu, C.~Zhang, and S.~Y. Philip, ``Direction-of-arrival estimation based
  on deep neural networks with robustness to array imperfections,'' {\em IEEE
  Transactions on Antennas and Propagation}, vol.~66, no.~12, pp.~7315--7327,
  2018.

\bibitem{crbStoicaNehorai}
P.~Stoica and A.~Nehorai, ``{MUSIC}, maximum likelihood, and {C}ram\'{e}r-{R}ao
  bound: {F}urther results and comparisons,'' {\em IEEE Transactions on
  Acoustics, Speech, and Signal Processing}, vol.~38, no.~12, pp.~2140--2150,
  1990.

\bibitem{friedlander}
B.~Friedlander and A.~Weiss, ``Direction finding in the presence of mutual
  coupling,'' {\em IEEE Transactions on Antennas and Propagation}, vol.~39,
  no.~3, pp.~273--284, 1991.

\bibitem{elbirSampTA}
A.~M. Elbir, S.~Mulleti, R.~Cohen, R.~Fu, and Y.~C. Eldar, ``Deep-sparse array
  cognitive radar,'' in {\em IEEE International Conference on Sampling Theory
  and Applications}, pp.~1--5, 2019.

\bibitem{performanceBoundsWWB}
A.~Renaux, P.~Forster, P.~Larzabal, C.~D. Richmond, and A.~Nehorai, ``A fresh
  look at the {B}ayesian bounds of the {Weiss-Weinstein} family,'' {\em IEEE
  Transactions on Signal Processing}, vol.~56, no.~11, pp.~5334--5352, 2008.

\bibitem{nion2010tensor}
D.~Nion and N.~D. Sidiropoulos, ``Tensor algebra and multidimensional harmonic
  retrieval in signal processing for {MIMO} radar,'' {\em IEEE Transactions on
  Signal Processing}, vol.~58, no.~11, pp.~5693--5705, 2010.

\bibitem{ye2008two}
Z.~Ye and C.~Liu, ``{2-D DOA} estimation in the presence of mutual coupling,''
  {\em IEEE Transactions on Antennas and Propagation}, vol.~56, no.~10,
  pp.~3150--3158, 2008.

\bibitem{quantizedCNN_Unified}
J.~{Cheng}, J.~{Wu}, C.~{Leng}, Y.~{Wang}, and Q.~{Hu}, ``Quantized {CNN}: {A}
  unified approach to accelerate and compress convolutional networks,'' {\em
  IEEE Transactions on Neural Networks and Learning Systems}, vol.~29, no.~10,
  pp.~4730--4743, 2018.

\bibitem{pan2010domain}
S.~J. Pan, I.~W. Tsang, J.~T. Kwok, and Q.~Yang, ``Domain adaptation via
  transfer component analysis,'' {\em IEEE Transactions on Neural Networks},
  vol.~22, no.~2, pp.~199--210, 2010.

\bibitem{freeezingLayers3}
J.~Yosinski, J.~Clune, Y.~Bengio, and H.~Lipson, ``How transferable are
  features in deep neural networks?,'' in {\em Advances in neural information
  processing systems}, pp.~3320--3328, 2014.

\bibitem{music}
R.~Schmidt, ``Multiple emitter location and signal parameter estimation,'' {\em
  IEEE Transactions on Antennas and Propagation}, vol.~34, no.~3, pp.~276--280,
  1986.

\bibitem{elbirMCM}
A.~M. {Elbir}, ``A novel data transformation approach for doa estimation with
  3-d antenna arrays in the presence of mutual coupling,'' {\em {IEEE} Antennas
  Wireless Propag. Lett.}, vol.~16, pp.~2118--2121, 2017.

\bibitem{vggRef}
K.~Simonyan and A.~Zisserman, ``Very deep convolutional networks for
  large-scale image recognition,'' {\em arXiv preprint arXiv:1409.1556}, 2014.

\bibitem{deepCNN_ChannelEstimation}
P.~{Dong}, H.~{Zhang}, G.~Y. {Li}, I.~S. {Gaspar}, and N.~{NaderiAlizadeh},
  ``{Deep CNN-Based Channel Estimation for mmWave Massive MIMO Systems},'' {\em
  {IEEE} J. Sel. Areas Commun.}, vol.~13, pp.~989--1000, Sep. 2019.

\bibitem{elbir2020DeepMUSIC}
A.~M. {Elbir}, ``{DeepMUSIC: Multiple Signal Classification via Deep
  Learning},'' {\em IEEE Sensors Letters}, vol.~4, no.~4, pp.~1--4, 2020.

\end{thebibliography}
	\begin{IEEEbiography}[{\includegraphics[width=1in,height=1.25in,clip,keepaspectratio]{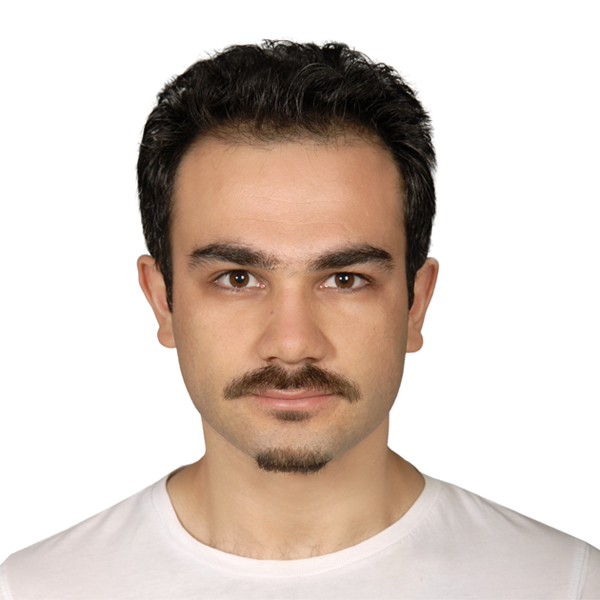}}]{Ahmet M. Elbir} [S'13--M'16--SM'20] received the B.S. degree with Honors from Firat University in 2009 and the Ph.D. degree from Middle East Technical University (METU) in 2016, both in electrical engineering. He is the recipient of 2016 METU best Ph.D. thesis award for his doctoral studies. He serves as an Associate Editor for IEEE Access since 2018. Currently, he is a visiting postdoctoral researcher at Koc University, Istanbul, Turkey. His research interests include array signal processing, sparsity-driven convex optimization, signal processing for communications and deep learning for array signal processing.
	\end{IEEEbiography}

	\begin{IEEEbiography}[{\includegraphics[width=1in,height=1.25in,clip,keepaspectratio]{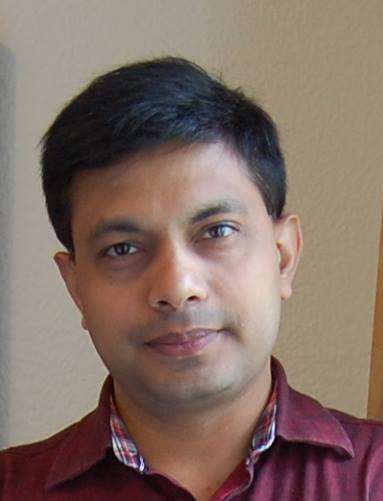}}]{Kumar Vijay Mishra} [S'08--M'15--SM'18] received B.Tech., summa cum laude (Gold medal, honors), in electronics and communications engineering from the National Institute of Technology, Hamirpur, India, in 2003, M.S. in electrical and computer engineering from Colorado State University, Fort Collins, in 2012, and his Ph.D. degree in electrical and computer engineering and M.S. degree in mathematics from The University of Iowa, Iowa City, in 2015 while working on NASA Global Precipitation Mission Ground Validation program weather radars. He is the recipient of IEEE MLSP Best Paper Award (2019), Royal Meteorological Society Quarterly Journal Editor's Prize (2017), Andrew and Erna Finci Viterbi Postdoctoral Fellowship (2015 and 2016), and Lady Davis Postdoctoral Fellowship (2016). Currently, he is visiting scholar at The University of Iowa; Research Fellow at the University of Luxembourg; and National Academies Harry Diamond Distinguished Fellow at the U. S. Army Research Laboratory. His research interests include remote sensing, signal processing, communications, deep learning, and electromagnetics.
	\end{IEEEbiography}
	
\end{document}